\documentclass[useAMS,usenatbib]{mn2e}
\usepackage{graphicx}
\usepackage{times}
\usepackage{amssymb}
\usepackage{natbib}

\newcommand{\be}{\begin{equation}}
\newcommand{\ee}{\end{equation}}

\def\ltsima{$\; \buildrel < \over \sim \;$}
\def\simlt{\lower.5ex\hbox{\ltsima}}
\def\gtsima{$\; \buildrel > \over \sim \;$}
\def\simgt{\lower.5ex\hbox{\gtsima}}

\renewcommand{\vec}[1]{ {\bmath #1} } 

\title{Secondary Infall and the Pseudo-Phase-Space Density Profiles of Cold Dark Matter Halos}

\author[Ludlow et al.] {\parbox{18cm}{
Aaron D. Ludlow$^{1,2},$ 
Julio F. Navarro$^{2}$,
Volker~Springel$^{3}$, 
Mark~Vogelsberger$^{3,5}$,\\ 
Jie~Wang$^{3,4}$,
Simon D.M. White$^{3}$, 
Adrian Jenkins$^{4}$, 
Carlos S. Frenk$^{4}$ 
}\vspace{0.3cm}\\
$^{1}${Argelander-Institut f\"{u}r Astronomie, Auf dem H\"{u}gel 71,
D-53121 Bonn, Germany}\\
$^{2}${Dept. of Physics and Astronomy, University of
    Victoria, Victoria, BC, V8P 5C2, Canada}\\
$^3$Max-Planck-Institut f\"{u}r Astrophysik,
Karl-Schwarzschild-Stra\ss{}e 1, 85740 Garching bei M\"{u}nchen,
Germany\\
$^{4}${Institute for Computational Cosmology, Dept. of Physics, Univ. of
  Durham, South Road, Durham  DH1 3LE, UK}\\
$^{5}${Harvard-Smithsonian Center for Astrophysics, 60 Garden Street,
  Cambridge, MA, 02138, USA}\\
}

\begin{document}

\maketitle 
\begin{abstract}
  We use N-body simulations to investigate the radial dependence of
  the density, $\rho$, and velocity dispersion, $\sigma$, in cold dark
  matter (CDM) halos. In particular, we explore how closely $Q\equiv
  \rho/\sigma^3$, a surrogate measure of the phase-space density,
  follows a power law in radius. Our study extends earlier work by
  considering, in addition to spherically-averaged profiles, local
  $Q$-estimates for individual particles, $Q_i$; profiles based on the
  ellipsoidal radius dictated by the triaxial structure of the halo,
  $Q_i(r')$; and by carefully removing substructures in order to focus
  on the profile of the smooth halo, $Q^s$. The resulting $Q_i^s(r')$
  profiles follow closely a power law near the center, but show a
  clear upturn from this trend near the virial radius, $r_{200}$. The
  location and magnitude of the deviations are in excellent agreement
  with the predictions from Bertschinger's spherical secondary-infall
  similarity solution. In this model, $Q \propto r^{-1.875}$ in the
  inner, virialized regions, but departures from a power-law occur
  near $r_{200}$ because of the proximity of this radius to the
  location of the first shell crossing---the shock radius in the case
  of a collisional fluid. Particles there have not yet fully
  virialized, and so $Q$ departs from the inner power-law profile.
  Our results imply that the power-law nature of $Q$ profiles only
  applies to the inner regions and cannot be used to predict
  accurately the structure of CDM halos beyond their characteristic
  scale radius.
\end{abstract}


\section{Introduction}
\label{sec:intro} 
\renewcommand{\thefootnote}{\fnsymbol{footnote}}
\footnotetext[1]{E-mail: aludlow@astro.uni-bonn.de}

The study of the clustering of cold dark matter (CDM) on the scale of
individual halos has progressed dramatically over the past couple of
decades due to the advent of powerful simulation techniques and ever
faster computers. As a result, a number of basic properties of the
structure of CDM halos are generally agreed upon, even when many of
these empirical findings lack a solid theoretical underpinning.  One
example is the approximately ``universal'' mass profile of virialized
CDM halos \citep[][hereafter NFW]{Navarro1996,Navarro1997}. CDM halos
are also strongly triaxial, with a preference for prolate shapes
\citep[see, e.g.,][and references
therein]{Frenk1988,Jing2002,Allgood2006,Hayashi2007}, and have
plentiful, albeit non-dominant, substructure
\citep[][]{Klypin1999,Moore1999}.

The mass profile of CDM halos can be well approximated by the simple
law proposed by NFW, where the logarithmic slope of the
spherically-averaged density profile follows the simple relation
$\gamma\equiv d\log{\rho}/d\log{r}=-(1+3x)/(1+x)$, with $x=r/r_{-2}$
being the radial coordinate expressed in units of a characteristic
halo scale radius, $r_{-2}$.  More recent work shows evidence for
small but systematic deviations from this simple law, and suggests
that a third parameter may actually be required to accurately describe
the mean mass profiles of CDM halos
\citep{Navarro2004,Merritt2005,Merritt2006,Gao2008,Hayashi2008}. These
authors argue that the density profile of $\Lambda$CDM halos steepens
monotonically with radius, with no sign of converging to a central asymptotic
power-law. The profiles are more accurately described by the
``Einasto'' formula for which $\gamma=-2(r/r_{-2})^{\alpha}$, with
$\alpha$ an adjustable ``shape'' parameter that can be tailored to
provide an improved fit to a given halo density profile.

As discussed by \citet[][]{Navarro2008}, this not only implies that
CDM halos are not strictly self-similar, but also makes it difficult
to predict the asymptotic properties of CDM halo mass profiles. For
example, the Einasto and NFW formulae predict quite different
asymptotic central behaviours: the Einasto profile has a true ``core''
with a well defined central density, whereas the central density of
the NFW profile diverges like $r^{-1}$. It seems clear from the latest
simulation results \citep[see, e.g.,][]{Navarro2008,Stadel2008} that
the asymptotic slope is shallower than $-1$, but it is unclear whether
the Einasto formula holds all the way to the centre and whether there
is truly a well-defined central density for CDM halos aside from the
ultimate upper bound set by the Tremaine-Gunn phase-space density
constraint \citep{TremaineGunn1979}. The gently curving nature of the
Einasto profile makes it difficult to extrapolate available
simulations in order to predict the central properties of the halo
with certainty.

One alternative is suggested by the realization that, although the
density, $\rho(r)$, and velocity dispersion, $\sigma(r)$ are complex
functions of radius, the quantity $Q(r)=\rho/\sigma^3$ follows closely
a simple power-law, $Q(r)\propto r^{\chi}$, with roughly the same
exponent, $\chi \sim -1.875$, for all halos \citep{Taylor2001}. $Q(r)$
has the same dimensions as the phase-space density, $f$, but it is not
a true measure of it \citep{Ascasibar2005,Sharma2006}. We shall
therefore refer to $Q$ as a pseudo-phase-space density, or as a
surrogate measure of phase-space density. Despite this, $Q$ relates
two moments of $f$ which occur often in equations that describe
equilibrium systems, and therefore simple relations between them are
extremely useful when constructing dynamical models \citep[see,
e.g.,][]{Austin2005,Dehnen2005,Barnes2006,Ascasibar2007}.

The proposal of \citet[][hereafter, TN]{Taylor2001} has been confirmed
in subsequent numerical work \citep[see,
e.g.,][]{Rasia2004,Dehnen2005,Faltenbacher2007,Vass2009,Wang2009} and
has been used in the literature to motivate dynamical models of dark
halos. One intriguing feature of the TN result is that the exponent of
the power-law $Q(r)$ profile is consistent with that found in the
similarity solutions of \citet[][]{Bertschinger1985}. Whether this is
a mere coincidence or has a deeper meaning remains unclear.  CDM halos
form through a combination of smooth infall and the accretion of
smaller progenitors that are subsequently disrupted in the tidal field
of the main halo. Bertschinger's similarity solution, on the other
hand, follows the accretion of radial mass shells onto a point-mass
perturber in an otherwise unperturbed Einstein-de Sitter universe. The
solution assumes spherical symmetry, allows only radial motions, and
is violently unstable \citep{Vogelsberger2009}. In spite of this, the
approximate power-law nature of $Q(r)$ has been confirmed by the latest series of
simulations, which resolve CDM halos with over one billion particles
\citep{Navarro2008}.

The actual value of the exponent has also received attention. Although
most simulations seem consistent with $\chi=-1.875$, best fits often
give slightly different values for $\chi$, typically in the range
$-1.85$ to $-2$ (but see \citealt{Schmidt2008} for a differing
view). Furthermore, there is an indication that $\chi$ may depend on the
main mode of mass accretion \citep{Wang2009}, and on the slope of the
primordial power-spectrum \citep{Knollmann2008}. The cited work {\it
  assumes} that $Q(r)$ is a power law, and then estimates $\chi$ from
simple fits to the spherically-averaged $Q$ profiles. If $Q$
profiles deviate slightly but significantly from a power law, it could
lead to a spread in the values of $\chi$, depending on, for example,
the radial range of the fits or the characteristic mass of the halos considered.

A further complication is introduced by the presence of
substructure. Although not dominant in mass, because of their higher
density and lower velocity dispersion, subhalos typically have much
higher values of $Q$ than the surrounding halo
\citep[][]{Arad2004,Diemand2008,Vass2009}. Together with the fact that
CDM halos are in general triaxial, this hinders a proper definition of
the value of $Q$ at given $r$, especially in the outer regions of a
halo, where subhalos are most abundant
\citep[][]{Springel2008b}.

We address these issues here using a series of high-resolution
cosmological N-body simulations of the formation of individual CDM
halos. In particular, we compute local estimates of $Q$ at each
particle position, $Q_i$, and contrast the resulting profiles with
those obtained using spherically-averaged estimates. The use of $Q_i$
allows us to carefully excise substructures and to focus our analysis on the
pseudo-phase-space density profile of the smooth main halo.

This paper is organized as follows. Sec.~\ref{SecNumSims} provides a
brief description of our numerical simulations; Sec.~\ref{SecProf}
discusses our main findings and compares them with Bertschinger's
similarity solutions. We conclude with a brief discussion of
our main conclusions in Sec.~\ref{SecSumm}.

\begin{figure*}
\begin{center}
\resizebox{18cm}{!}{\includegraphics{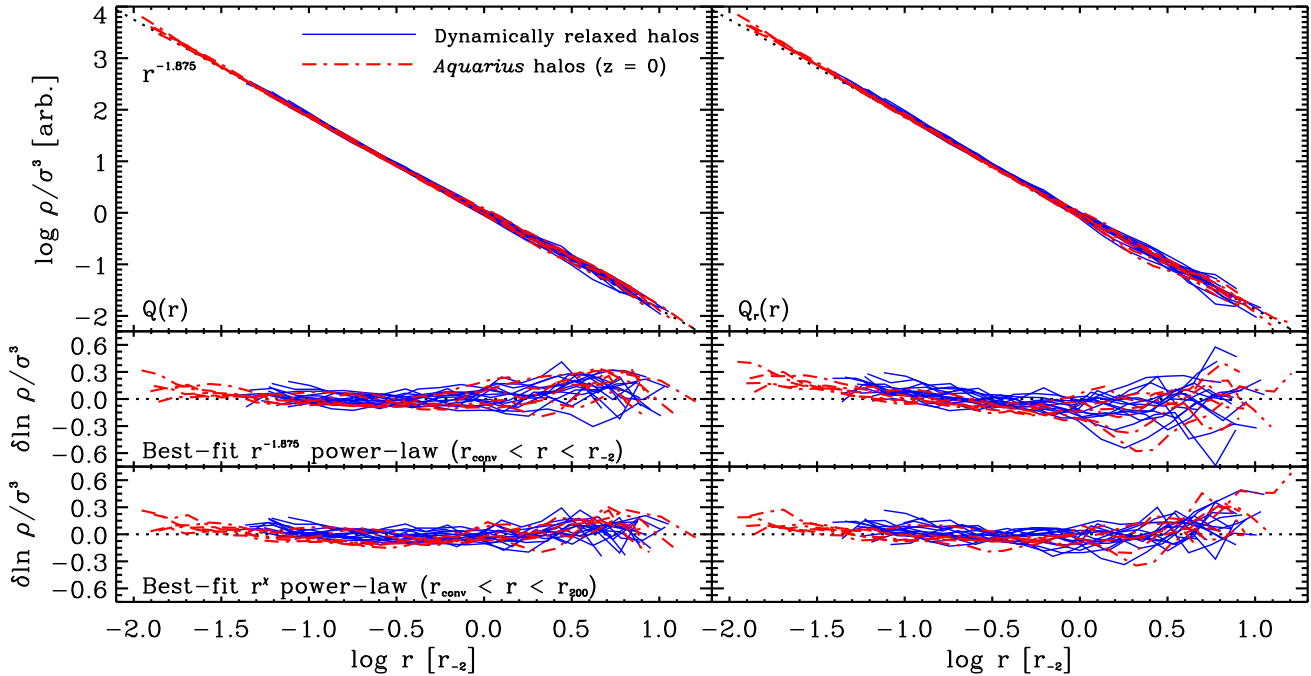}}
\end{center}
\caption{Spherically averaged pseudo-phase-space density profiles for
  the 21 dark matter halos in our sample. The six level-2 Aquarius
  halos are shown at $z=0$ (red dot-dashed lines), the other fifteen
  (solid blue) are shown at the most recent redshift when they pass
  the dynamical relaxation criteria (Sec.~\ref{SecRelCrit}). Left
  panels correspond to $Q(r)\equiv \rho/\sigma^3$; panels on the right
  correspond to the ``radial'' $Q_r(r)\equiv \rho/\sigma_r^3$. Radii
  are scaled to the scale radius, $r_{-2}$, of each halo. Middle
  panels show residuals from the best $r^{-1.875}$ power-law fit to
  the $r_{\rm conv}<r<r_{-2}$ portion of the profiles. These best fits
  are also used to choose the vertical normalization of each profile
  in the upper panels, so as to minimize the halo-to-halo scatter in
  the inner profiles. Bottom panels are analogous to the middle ones,
  but for power-law fits over the whole range $r_{\rm
    conv}<r<r_{200}$, with free-floating exponent, $r^{\chi}$.
  Values of $\chi$ and $\chi_r$ for each halo are listed in
  Table~\ref{TabChiFits}.}
\label{FigSphQProf}
\end{figure*}

\section{The Simulations}
\label{SecNumSims}

We study the formation of $21$ CDM halos selected from a $100\,
h^{-1}$ Mpc-box cosmological simulation and resimulated at high
resolution in their full cosmological context. We provide below a
brief summary of the numerical techniques, including the adopted
cosmological parameters; the initial conditions setup; the simulation
code; as well as the halo selection criteria and analysis
techniques. More detailed information about our resimulation and
analysis techniques may be found in previous papers by our group
\cite[e.g.,][]{Power2003,Navarro2004,Springel2008b,Navarro2008}.

\subsection{Cosmological Parameters}
\label{SsecCosmParam}

All our simulations adopt the currently favored $\Lambda$CDM cosmogony
with the following parameters: $\Omega_{\rm M}=0.25$,
$\Omega_{\Lambda}=1-\Omega_{\rm M}=0.75$, $\sigma_8=0.9$, $n_s = 1$,
and a Hubble constant $H_0=H(z=0)=100\, h$ km s$^{-1}$ Mpc$^{-1}=73$
km s$^{-1}$ Mpc$^{-1}$. These parameters are the same as those adopted
for the Millennium Simulation \citep{Springel2005a}, and are
consistent (within 2-sigma) with those derived from the WMAP 1- and 5-year data
analyses \citep[]{Spergel2003,Komatsu2009} and with the recent cluster
abundance analysis of \citet[]{Henry2009}.

\subsection{Halo Selection}
\label{SsecHaloSel}

The $21$ halos were selected from the same $900^3$-particle, $100 \,
h^{-1}$ Mpc-box parent simulation used for the Aquarius Project
\citep{Springel2008b}. These halos were subsequently resimulated at
higher resolution using the technique described in detail by
\citet{Power2003}. We avoid halos that form in the periphery of much
larger systems by imposing a mild isolation criterion so that no halo
more massive than half the mass of the selected system lies within
$1\, h^{-1}$ Mpc at $z=0$. This parent simulation was later also
resimulated in its entirety at much higher resolution; this is the
Millennium-II Simulation recently analyzed by
\citet{Boylan-Kolchin2009}.

Besides the six {\it Aquarius} halos, which were all selected to have virial
\footnote{We define the virial mass of a halo, $M_{200}$, as that
  contained within a sphere of mean density $200$ times the critical
  density for closure, $\rho_{\rm crit}=3H^2/8\pi G$. The virial mass
  defines implicitly the virial radius, $r_{200}$, and virial
  velocity, $V_{200}=(GM_{200}/r_{200})^{1/2}$, of a halo,
  respectively.}
masses of order $\sim 10^{12} \, h^{-1} M_{\odot}$, we have
resimulated a further set of $15$ halos in order to span the mass
range $\sim 10^{12}$ to a few times $10^{14}\, h^{-1}
M_{\odot}$. These $15$ simulations have typically a few million
particles within the virial radius at $z=0$ and are of lower numerical
resolution than the level-2 Aquarius halos. Combining these two 
datasets allows us to assess the sensitivity of our results to 
numerical resolution. Table~\ref{TabNumSims} lists the main properties 
of each halo in our sample.

\subsection{The Code}
\label{SsecCode}

All simulations were run with either the publicly available GADGET-2
code \citep{Springel2005b} or its latest version, GADGET-3, which was
developed for the Aquarius Project. Softening
lengths are chosen according to the ``optimal'' prescription of
\citet{Power2003}. Pairwise interactions are fully Newtonian for
separations exceeding the spline-lengthscale
$h_s$. Table~\ref{TabNumSims} quotes the equivalent Plummer softening,
$\epsilon_G=h_s/2.8$, for each resimulated halo. Throughout the
simulations, the softening length is kept fixed in comoving coordinates.

\subsection{Analysis}
\label{SsecAnal}

\subsubsection{Spherically-averaged $Q$ profiles}

In order to compute the spherically-averaged pseudo-phase-space
density profile of each halo we first identify the halo center with
the location of the particle having the minimum potential energy. Then
we compute $Q(r)$ in $N_{\rm bin}$ spherical shells equally spaced in
$\log_{10} r$ in the range $r_{\rm conv}\leq r\leq r_{200}$. Here
$r_{\rm conv}$ is the convergence radius defined by \citet{Power2003},
where circular velocities converge to better than $10\%$ \citep[see
also][]{Navarro2008}. For each spherical shell (radial bin), we
estimate $Q(r)=\rho/\sigma^3$, where $\rho$ is just the mass of the
shell divided by its volume and $\sigma^2$ is twice
the specific kinetic energy in the shell. We also
compute a ``radial'' $Q$ estimate, $Q_r(r)=\rho/\sigma_r^3$ in an
analogous way, although instead of the total kinetic energy we use
only the kinetic energy in radial motions to estimate $\sigma_r$.

\begin{figure}
\begin{center}
\resizebox{8.5cm}{!}{\includegraphics{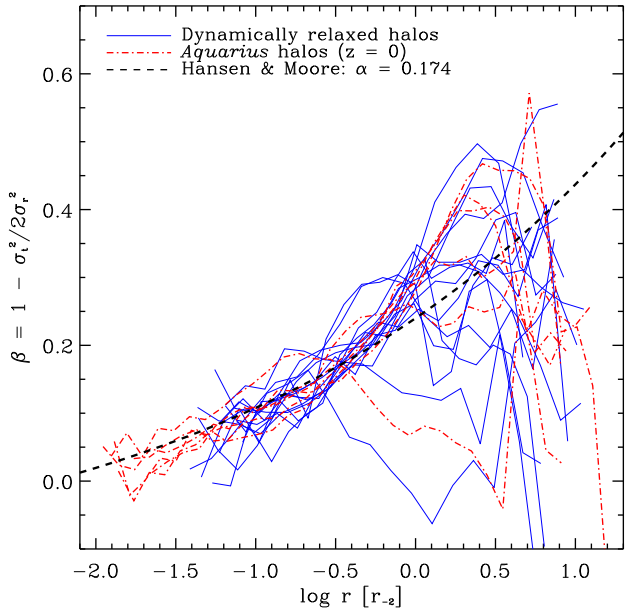}}
\end{center}
\caption{Radial profiles of the velocity anisotropy parameter,
  $\beta=1-\sigma_t^2/2\sigma_r^2$, for all halos in our
  sample. Line types are as in Fig.~\ref{FigSphQProf}. Radii are
  normalized to the scale radius of each halo. Note the non-monotonic
  radial dependence of $\beta$: halos are nearly isotropic near the
  center, radially biased near $r_{-2}$, but approximately isotropic
  again in the outskirts. The anisotropy expected from the
  $\beta(\gamma)$ relation of \citet{Hansen2006} for an Einasto
  profile with $\alpha=0.174$ is also shown. 
}
\label{FigAnisProf}
\end{figure}

\begin{figure}
\begin{center}
\resizebox{8.5cm}{!}{\includegraphics{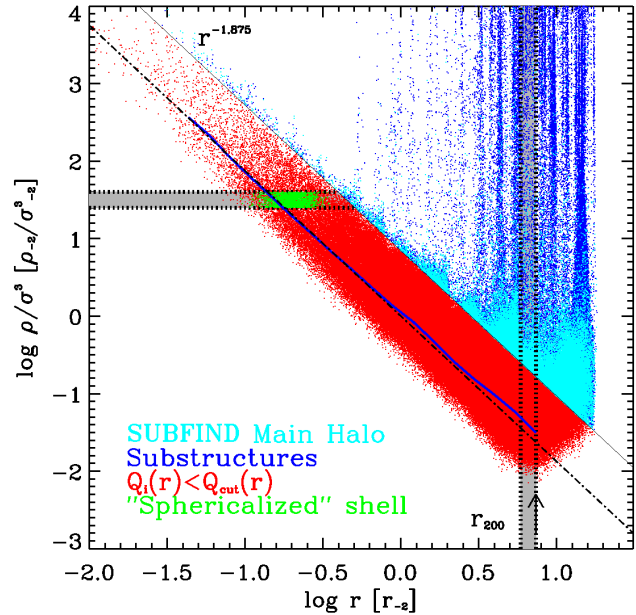}}
\end{center}
\caption{``Local'' estimates of the pseudo-phase-space density, $Q_i$,
  as a function of distance from the halo centre for all halo particles in
  one of our simulations. Different colors correspond to various
  particle subsamples. (i) Blue denotes particles in self-bound
  substructures, as identified by {\small SUBFIND}. (ii) Cyan
  indicates particles not bound to any substructure but with
  higher-than-average $Q_i$ for their location in the halo. These are
  particles in recently-stripped tidal streams which, although
  assigned to the main halo by {\small SUBFIND}, have yet to phase mix
  within the main halo \citep{Maciejewski2009}.  (iii) Red dots
  indicate particles with $Q_i<Q_{\rm cut}(r)$, which we define as
  belonging to the relaxed main halo (see Fig.~\ref{FigQDist}). Also
  plotted are the spherically-averaged $Q(r)$ profile (solid blue
  line), and the best-fit $r^{-1.875}$ power law (dot-dashed
  line). Vertical and horizontal bands show, respectively, the
  particles selected for Figs.~\ref{FigQDist} and ~\ref{FigQShell}.}
\label{FigQiProf}
\end{figure}

\subsubsection{Local $Q$ profiles}

A different estimate of the pseudo-phase-space density may be
obtained, for each shell, by considering ``local'' estimates of $Q$ at
the position of each particle. We shall call this
$Q_i=\rho_i/\sigma_i^3$, and use $N_{\rm ngb}$ nearest neighbours in
order to compute the local density and velocity dispersion. Density
estimates at the location of the $i^{th}$ particle are computed as
\begin{equation}
\rho_i=\sum_{j=1}^{N}m_j W(|\vec{r}_{i,j}|,h_i),
\end{equation}
where $\vec{r}_{i,j}\equiv\vec{r}_i-\vec{r}_j$, and $W(r,h)$ is the
smoothing kernel often adopted in Smoothed Particle Hydrodynamics
(SPH) simulations:
\begin{equation} 
W(r,h)=\frac{8}{\pi h^3} \left\{
\begin{array}{ll}
1-6\left(\frac{r}{h}\right)^2 + 6\left(\frac{r}{h}\right)^3, &
0\le\frac{r}{h}\le\frac{1}{2} ,\\
2\left(1-\frac{r}{h}\right)^3, & \frac{1}{2}<\frac{r}{h}\le 1 ,\\
0 , & \frac{r}{h}>1 .
\end{array}
\right.
\end{equation}
The smoothing length, $h_i$, of each particle is defined implicitly
by the smallest volume that contains $N_{\rm ngb}$ nearest neighbours:
\begin{equation}
h_i = \biggl( N_{\rm ngb}\frac{3}{4\pi}\frac{m_i}{\rho_i}\biggr)^{1/3}.
\end{equation}
We use, as default, $N_{\rm ngb}=48$ for our lower resolution runs and $64$ for the
level-2 Aquarius halos. 

Given $N_{\rm ngb}$, the local velocity dispersion for particle $i$ is
given by $\sigma_{i}^2=\overline{v^2}-\overline{v}^2$, where the
unweighted averages are computed over all $N_{\rm ngb}$ neighbours.

\subsubsection{Relaxation criteria}
\label{SecRelCrit}

In order to minimize the effect of transient, rapidly-evolving
evolutionary stages, such as ongoing mergers, we impose (when
explicitly stated) relaxation criteria similar to those introduced by
\citet{Neto2007}. These include restrictions on the fraction of the
virial mass in self-bound substructures, $f_{\rm sub}=M_{\rm
  sub}(r<r_{200})/M_{200}<0.07$; on the offset between the center of
mass of the halo and its true centre (as defined by the particle with
minimum potential energy), $d_{\rm off} = |\vec{r}_{\rm
  CM}-\vec{r}_{\rm cen}|/r_{200}<0.05$, and on the virial ratio of
kinetic to potential energies, $2K/|\Phi|<1.3$.

In practice, when a halo does not satisfy the relaxation criteria at
$z=0$ we track its main progenitor back in time until we find the
first snapshot when it does. This typically occurs at redshifts less
than $\sim 0.2$, but in one case we had to go back in time until
$z\sim 0.8$ in order to find a suitably ``relaxed'' configuration. In
what follows, we shall consider relaxed configurations only for the
lower-resolution halos but take the $z=0$ configuration for the
Aquarius halos. As we show below, the results are similar in the two
cases, which means that our conclusions are not particularly sensitive
to our requirement of dynamical equilibrium.

The properties of each halo in our sample at $z=0$ are listed in
Table~\ref{TabNumSims}. Here we list the virial mass, $M_{200}$, the
virial radius, $r_{200}$, the number of particles ($N_{200}$) within
$r_{200}$, as well as the gravitational softening, $\epsilon_G$, and
the convergence radius, $r_{\rm conv}$. The peak of the circular
velocity curve is also specified by $r_{\rm max}$ and $V_{\rm
  max}$. 

\section{Pseudo-Phase-Space Density Profiles}
\label{SecProf}

\begin{figure}
\begin{center}
\resizebox{8.5cm}{!}{\includegraphics{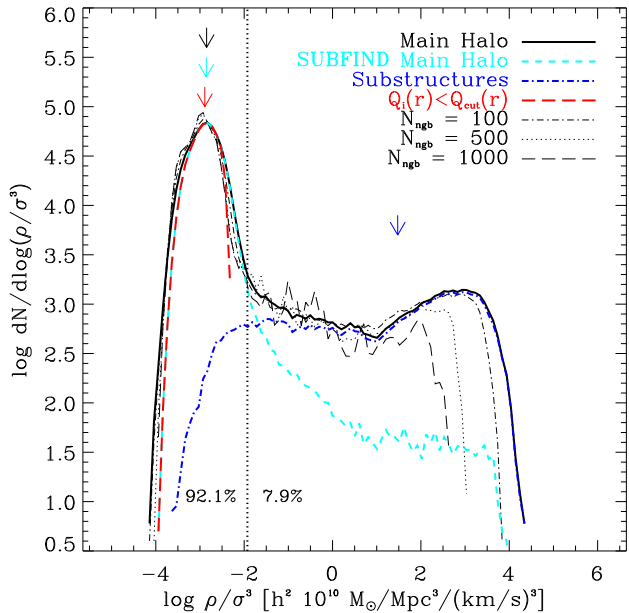}}
\end{center}
\caption{Distribution of local phase-space densities, $Q_i$, for
  particles in the thin (shaded) spherical shell near the virial
  radius of the halo shown in Fig.~\ref{FigQiProf}. Substructures
  identified by {\small SUBFIND} are shown in blue, the main {\small
    SUBFIND} halo in cyan. Various thin lines illustrate the effect
  of varying the number of neighbours in the $\rho_i$ and $\sigma_i$
  estimates, as labelled in the legend. As discussed in the text, a
  simple cut $Q_i<Q_{\rm cut}$ identifies unequivocally all well-mixed
  particles in the main halo; $Q_{\rm cut}$ is shown by the vertical
  dotted line. Down-pointing arrows indicate the median of the
  distribution of each set of particles. Because there are few
  particles in the high-$Q_i$ tail, the median $Q_i$ of the main
  {\small SUBFIND} halo and that of particles with $Q_i<Q_{\rm cut}$
  are nearly identical. We adopt the latter as the characteristic
  pseudo-phase-space density of the smooth halo at each radius.}
\label{FigQDist}
\end{figure}

\subsection{Spherically-averaged Q profiles}
\label{SecSphAvProf}

Figure~\ref{FigSphQProf} shows the spherically-averaged $Q(r)$
profiles for all halos in our sample, together with residuals from
various best-fits. Panels on the left and on the right correspond to
$Q(r)$ and $Q_r(r)$, respectively. The plotted profiles extend from
the convergence radius, $r_{\rm conv}$, to the virial radius,
$r_{200}$.  Middle panels show residuals from best fits to the
region inside $r_{-2}$ with an $r^{-1.875}$ power law. All profiles are
normalized to the scale radius, $r_{-2}$, and vertically according to
the power-law best fit. Bottom panels show residuals from fits to the
$r_{\rm conv}<r<r_{200}$ profile with a power law with free-floating
exponent, $Q\propto r^{\chi}$.

A few things are worth noting in this figure. The first is how closely
both the $Q(r)$ and $Q_r(r)$ profiles follow simple power laws, from
the innermost resolved radius out to $r_{200}$. In the case of $Q(r)$,
even when the exponent of the fit is fixed at $\chi=-1.875$, which
means that a {\it single} free parameter (the vertical normalization)
is allowed, residuals from best fits do not exceed $\sim 30\%$ {\it
  anywhere} within the virial radius. This power-law behaviour holds
for roughly three decades in $r$ and six decades in $Q$. 

Although the best-fit $\chi$ differs from $-1.875$ (see
Table~\ref{TabChiFits} for actual values), the residuals decrease only
very slightly when allowing $\chi$ to float freely. Defining a
figure-of-merit function as
\begin{equation}
  \psi^2= {1\over N_{\rm bins}} \sum_1^{N_{\rm bins}}
  (\ln Q - \ln Q_{\rm fit})^2, 
\label{EqFigofMerit}
 \end{equation}
 we find that $\psi$, averaged over all halos and fit over the range
 $r_{\rm conv}<r<r_{200}$, varies only from $\langle \psi \rangle
 =0.102$ when fixing $\chi=-1.875$ to $\langle \psi \rangle =0.085$
 when allowing $\chi$ to be a free parameter.

The ``tilt'' in the $Q_r$-residuals shown in the middle-right panel of
Figure~\ref{FigSphQProf} indicates that the $\chi_r$ exponent that fits
best the $Q_r(r)$ profiles is slightly more negative than
$-1.875$. Overall, however, $Q_r(r)$ may also be approximated with a
power law with $\chi_r\approx 1.92$, as may be seen in the bottom right
panel of the same figure.  The average $\psi$ for power-law fits with
variable $\chi_r$ is $0.142$, which means that $Q_r(r)$ deviates
 more than $Q(r)$ from a simple power law, but only slightly so.

 The reason why $Q_r(r)$ deviates more from a simple power law than
 $Q(r)$ may be traced to the complex behaviour of the anisotropy
 parameter, $\beta(r)=1-\sigma_{\rm t}^2/2\sigma_r^2$. All halos in
 our sample are almost isotropic near the center, radially anisotropic
 further out, but nearly isotropic again close to the virial
 radius. Because of this complex radial behaviour, $Q(r)$ and $Q_r(r)$
 cannot both be simultaneously well fitted by a power law with the
 same exponent.

Although power laws provide excellent fits to the
pseudo-phase-space density profiles, further scrutiny of the middle
and lower panels of Fig.~\ref{FigSphQProf} reveals a well defined
trend in the residuals of most halos, which tend to ``curve up''
slightly but significantly in the outer regions and, to a lesser
extent, in the innermost regions as well.  The latter deviations are
best appreciated in the Aquarius halos, which have much better
resolution than the rest.

These results seem to apply both to the Aquarius halos and to the
dynamically-relaxed halo sample, which suggests that our conclusions
are not crucially dependent on the adoption of the particular
relaxation criteria we used to select the sample. The above-noted
trend in the residuals means that the exponent, $\chi$, derived from
$r^{-\chi}$ power-law fits will depend on the radial range adopted for
the fit.

Given the desirable properties of a simple power law, it is worth
investigating whether the deviations from a simple $r^{\chi}$
behaviour might be due to the presence of substructure or to the
aspherical nature of halo structure. We explore these possibilities
next.

\begin{figure*}
\begin{center}
\resizebox{16cm}{!}{\includegraphics{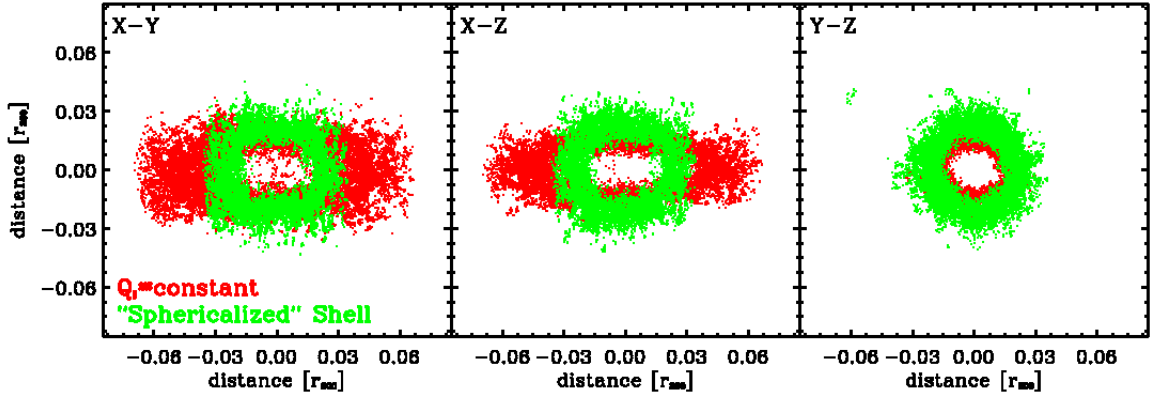}}
\end{center}
\caption{Orthogonal projections of particles in a shell of roughly
  constant pseudo-phase-space density (i.e., those in the horizontal
  shaded band in Figure~\ref{FigQiProf}, excluding substructures). For
  clarity, we plot in each panel only particles in a thin slice
  perpendicular to the viewing axis. Red points show the original
  positions; these delineate a nearly prolate ellipsoid, which has
  been rotated so that its principal axes coincide with the coordinate
  axes of the plot. Green points show the projected particle positions
  after correcting for triaxiality by the method outlined in the
  text.}
\label{FigQShell}
\end{figure*}

\subsection{Local Q profiles}
\label{SecQiProf}

Fig.~\ref{FigQiProf} shows $Q_i$, the local estimate of $Q$ at the
location of each particle in the halo, as a function of the distance
from the halo centre. Because substructures are overdense and have
lower velocity dispersion than their immediate surroundings, they show
up prominently in this plot as particles with very high $Q_i$ at a
given radius. This is confirmed by the color coding adopted in the
figure: particles in dark blue are those associated by the
substructure-finder {\small SUBFIND} \citep{Springel2001b} to
self-bound subhalos that survive within the main halo. Clearly,
because their pseudo-phase-space density is so distinct from the main
halo, substructures have the potential to bias estimates of $Q(r)$,
especially in the outer regions, where subhalos are more prevalent.

Although it would be simple enough to remove the self-bound structures
from $Q(r)$ profiles, the cyan dots in Fig.~\ref{FigQiProf}
illustrate a second, related problem. These are particles that SUBFIND
associates with the main halo, but which clearly have deviant $Q_i$
values relative to the surrounding average. As discussed, for example,
by \citet{Maciejewski2009}, these are particles recently stripped from
substructures; although now unbound to any subhalo, they have yet to
phase mix fully with the underlying main halo.

Fig.~\ref{FigQDist} shows the $Q_i$ distribution of all particles in
the thin spherical shell near the virial radius of the halo shown by
the vertical shaded region in Fig.~\ref{FigQiProf}. This shows the
wide range in $Q_i$ (almost 8 decades) spanned by particles at a fixed
distance from the halo centre. Despite this, the figure also shows
that the characteristic $Q$ at that distance is well defined, since
most particles have $Q_i$ values close to the peak on the left of the
distribution. (Note the logarithmic scale in the $y$-axis.) The tail
to the right of this peak contains self-bound substructures (in blue,
as identified by {\small SUBFIND}) and recently stripped material,
which, as mentioned above, are assigned to the main halo by {\small
  SUBFIND} despite their higher-than-average $Q_i$ values.

The shape of the distribution suggests that imposing a simple
criterion, e.g., $Q_i<Q_{\rm cut}$, should be sufficient to identify
unequivocally well-mixed particles belonging to the main smooth halo;
a plausible choice for $Q_{\rm cut}$ is shown by the vertical dotted
line in Fig.~\ref{FigQDist}. Down-pointing arrows indicate the median
of the distribution of each set of particles. Note that, because a
small fraction of the particles are in the high-$Q_i$ tail, the median
$Q_i$ of the main {\small SUBFIND} halo and that of particles with
$Q_i<Q_{\rm cut}$ are nearly identical. 

We therefore adopt the simple $Q_i<Q_{\rm cut}$ prescription to define
the main ``smooth'' halo. Once $Q_{\rm cut}$ is chosen at some radius,
we may use the approximate power-law behaviour of $Q$ to scale it to
any other radius by $Q_{\rm cut}(r) \propto r^{-1.875}$. Particles
shown in red in Fig.~\ref{FigQiProf} are those assigned to the smooth
halo by this criterion. At each radius, we shall adopt the median
$Q_i$ of these particles in order to construct the pseudo-phase-space
density profile of the main smooth halo. We shall refer to this
profile as $Q_i^s(r)$.

We note that this definition is insensitive to the number of nearest
neighbors ($N_{\rm ngb}$) adopted to compute $Q_i$: the various lines
in Fig.~\ref{FigQDist} illustrate the results for $N_{\rm ngb}=48$
particles (our default value), as well as for $1000$ (long-dashed),
500 (dotted), and 100 (dot-dashed), respectively.

\begin{figure}
\begin{center}
\resizebox{8.5cm}{!}{\includegraphics{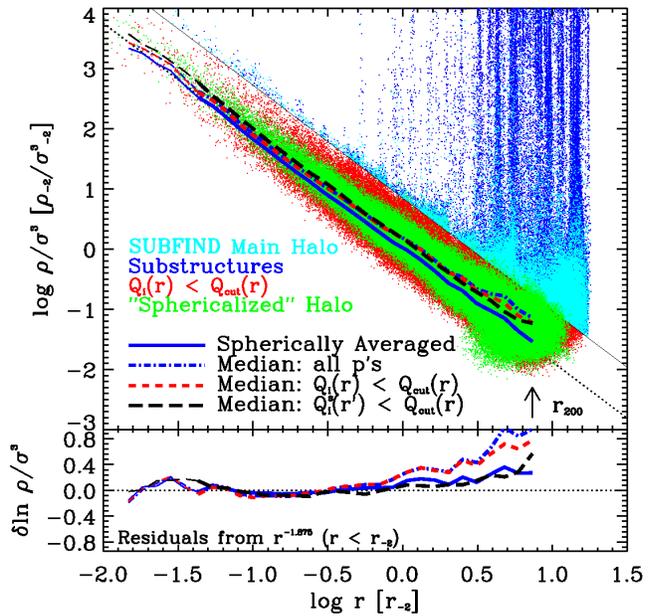}}
\end{center}
\caption{Local pseudo-phase-space density as a function of radius for
  the same halo shown in Fig.~\ref{FigQiProf}. Colored dots are as in
  Fig.~\ref{FigQiProf}. Red and green dots indicate smooth main halo
  particles before and after correcting for halo shape. The solid
  curve is the spherically-averaged profile, and is compared with the
  curves tracing the median $Q_i$ for (i) all particles (blue
  dot-dashed), (ii) all particles in the smooth main halo (red
  dashed), and (iii) smooth main halo particles corrected for
  triaxiality (dashed). The dotted line shows a power-law,
  $r^{-1.875}$.}
\label{FigSphQiProf}
\end{figure}

\begin{figure*}
\begin{center}
\resizebox{16cm}{!}{\includegraphics{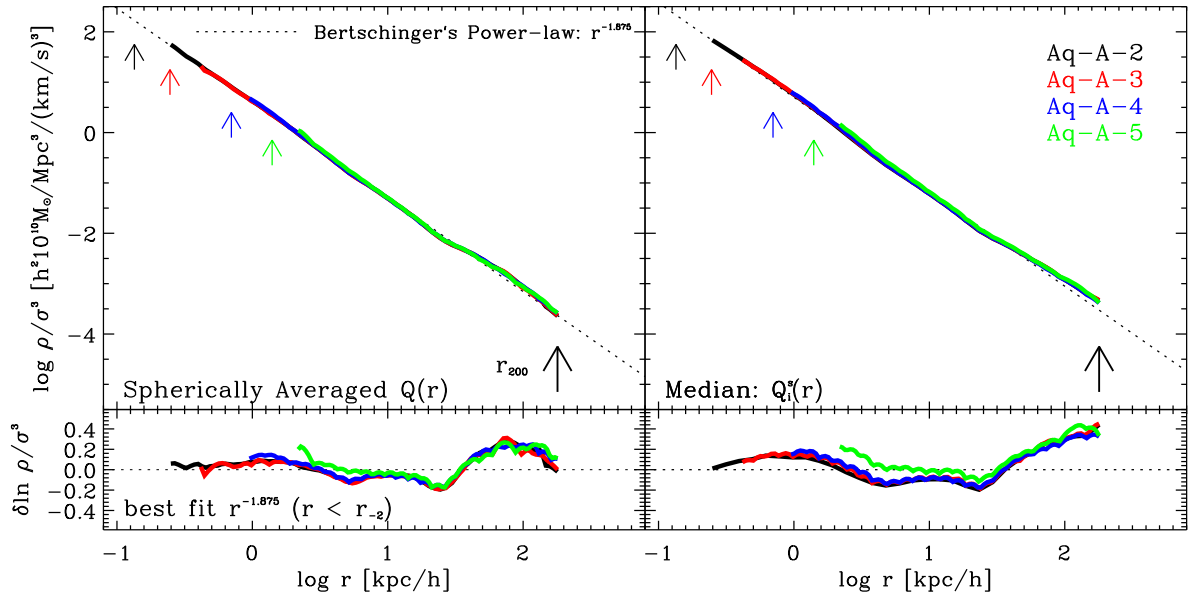}}
\end{center}
\caption{Spherically-averaged (left) and median local (right)
  pseudo-phase-space density profiles for the convergence
  series of the Aq-A halo \citep{Springel2008b,Navarro2008}. Different colors correspond to runs
  with different resolution, as indicated in the legend. Local
  profiles refer to the median $Q_i$ of the smooth main halo as a
  function of distance to the halo centre. The dotted line is the
  power-law $Q\propto r^{-1.875}$, scaled to match Aq-A-2 at $r\leq
  r_{-2}=11.15 \, h^{-1}$ kpc. Bottom panels show the residual with respect to this
  power-law.}
\label{FigNumConv}
\end{figure*}

\subsection{Correction for triaxiality}
\label{SecTriax}

Dark halos are not spherically symmetric. Because of this, a shell of
particles at constant distance from the halo center will have a wide
$Q_i$ distribution, even if one subtracts substructure as specified in
the previous subsection. Iso-$Q$ surfaces track fairly well the
isodensity contours of the main halo, and follow closely
three-dimensional ellipsoidal surfaces.  Fig.~\ref{FigQShell} shows
three orthogonal projections of particles with similar values of
$Q_i$, selected from those falling in the horizontal band highlighted
in Fig.~\ref{FigQiProf}. Only particles in the smooth main halo are
plotted here. The original particle positions (in red) in a thin slice
perpendicular to the line of sight are seen to trace a nearly prolate
ellipsoid, which for convenience has been rotated so that its
principal axes coincide with the coordinate axes of the projection.

Given that the ``iso-$Q$'' surfaces are well approximated by
ellipsoids, we may use the eigenvalues of the diagonalized inertia
tensor to compute an elliptical radius, $r'$, for each $Q_i$, to
define an ``ellipsoidal'' $Q_i(r')$ profile that may be contrasted
directly with $Q_i(r)$. In practice, we slice the smooth main halo in
narrow bins in $Q_i$; compute the axis lengths $a$, $b$, and $c$; and
use them to reassign an ellipsoidal radius $r'$ to each particle in
the smooth main halo.  We compute $r'$ as
\begin{equation}
r'^2=\biggr( \frac{x}{a}\biggl)^2 + \biggr( \frac{y}{b}\biggl)^2 + 
\biggr( \frac{z}{c}\biggl)^2,
\label{EqRell}
\end{equation}
after normalizing $a$, $b$, and $c$ so that $abc=1$ for all
shells. This choice preserves the typical distance to the halo centre
of particles in a given $Q_i$ shell. The result of the procedure is
illustrated in Fig.~\ref{FigQShell}, where the green dots are the same
particles as those shown in red, but the coordinate axes are now
$x'=x/a$, $y'=y/b$ and $z'=z/c$.

The green dots in Fig.~\ref{FigSphQiProf} delineate the $Q_i^s(r')$
profile for the smooth main halo.  The lines in
Fig.~\ref{FigSphQiProf} show the variations in the pseudo-phase-space
density profile induced by the various alternative ways of estimating
distances and $Q$ that we have discussed so far.

Although the profiles change appreciably, they all show the same
upturn in the outer regions, relative to a simple $r^{-1.875}$
power-law fit, noted when discussing the spherically-averaged $Q(r)$
profiles (Sec.~\ref{SecSphAvProf}). The upturn is indeed more
pronounced when using local-$Q$ estimates, because (i) local densities
are sensitive to inhomogeneities and are therefore higher than the
spherical average, and (ii) because local $\sigma$ estimates are lower
than the spherical average, since they do not include the bulk motion
of subhalos and recently stripped material. Interestingly,
Fig. ~\ref{FigSphQiProf} shows that, after correcting for triaxiality,
the spherically-averaged profile is indistinguishable from the
$Q_i^s(r')$ profile. We conclude that the upturn is {\it not} caused
by the presence of substructures in the outer regions nor by
departures from spherical symmetry. We discuss the interpretation of
this robust feature of the $Q$ profiles next.

\subsection{Numerical convergence}
\label{SecNumfConv}

Before considering the meaning of the departures of $Q$ profiles from
simple power laws we should check explicitly that our conclusions are
not affected by the numerical resolution of the simulations.  We show
this in Fig.~\ref{FigNumConv}, where we compare the $Q(r)$ and
$Q_i^s(r)$ profiles of one of the Aquarius halos at four different
resolutions. The highest, Aq-A-2, has more than $100$ million
particles within the virial radius; the lowest, Aq-A-5, about $600$
thousand. The non-Aquarius halos in the series analyzed in this paper
have numerical resolution comparable to Aq-A-4.
Fig.~\ref{FigNumConv} shows convincingly that our results are unlikely
to be adversely affected by numerical resolution. Both the
spherically-averaged and the local pseudo-phase-space density profiles
are very well reproduced at all radii, down to the inermost resolved
radius, $r_{\rm conv}$, of each run.

\subsection{Comparison with Bertschinger's similarity solution}
\label{ssec:SSIMcomp}

\begin{figure}
\begin{center}
\resizebox{8.5cm}{!}{\includegraphics{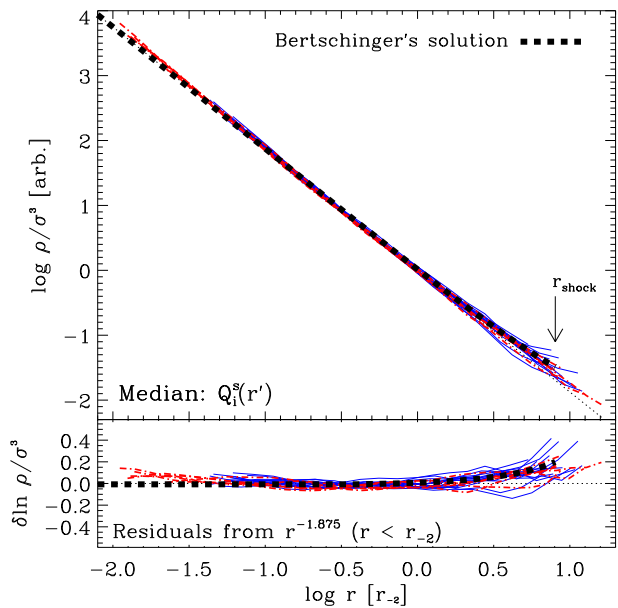}}
\end{center}
\caption{Local estimates of the pseudo-phase-space density as a
  function of the ellipsoidal radius $r'$ (eq.~\ref{EqRell}). Each
  curve traces the median $Q_i^s(r')$ for smooth main halo particles,
  computed in $r'$ bins of equal logarithmic width. Radii are scaled
  to the value of $r_{-2}$ of each halo. The dotted line shows the
  $Q\propto r^{-1.875}$ power law, whereas the thick dashed black curve shows
  the actual similarity solution computed by
  \citet{Bertschinger1985}. The outermost point of this solution,
  indicated by the downward pointing arrow, corresponds to $r_{\rm shock}$, the ``shock radius'' (for
  collisional fluid infall) or to the position of the first caustic
  (for collisionless fluids). Radii for the similarity solution assume
  $r_{\rm shock}=8\, r_{-2}$ All profiles are normalized at
  $r'=r_{-2}$. Note that halo profiles deviate from the asymptotic
  inner $r^{-1.875}$ power law in the same way as the similarity
  solution.}
\label{FigSimSolProf}
\end{figure}

The local pseudo-phase-space density profiles for the smooth main halo
of all our systems is shown in Fig.~\ref{FigSimSolProf}. The profiles
are shown as a function of the elliptical radius (eq.~\ref{EqRell}),
scaled to the scale radius of each halo, $r_{-2}$. All profiles have
been normalized vertically so that they coincide at $r'=r_{-2}$. The
dotted line shows a $Q \propto r^{-1.875}$ power law, also
normalized at $r_{-2}$. The bottom panel shows residuals relative to
the $r^{-1.875}$ power law. 

Fig.~\ref{FigSimSolProf} shows the main result of our analysis. The
pseudo-phase-space density profiles of our simulated halos clearly
deviate from a simple power law in the outer regions. This deviation
is actually {\it predicted} by the secondary infall similarity
solution of \citet{Bertschinger1985}. Indeed, in the similarity
solution the $Q\propto r^{-1.875}$ behaviour occurs only
asymptotically in the inner regions of the halo. In the outer regions,
and more precisely, near the location of the ``shock radius'', $r_{\rm
  shock}$ (for collisional fluids), or, equivalently, of the first
infall caustic (for collisionless fluids), the pseudo-phase-space
density profile (or entropy profile in the case of a collisional
fluid) shows a clear upturn from the inner asymptotic power-law
behaviour.

The reason for this upturn is that the fluid is not fully virialized
near $r_{\rm shock}$, since this radius marks the transition between
mass shells that are infalling for the first time and those that have
already crossed (or shocked) material that collapsed earlier. For
example, in the case of collisionless fluids, a mass shell must cross
the center and complete roughly 2-3 full oscillations before settling
onto a periodic orbit of constant apocenter. In the case of a
collisional fluid, a newly shocked shell drifts inward from the radius
at which it was shocked before reaching hydrostatic equilibrium (see
Fig.~4 of \citealt{Bertschinger1985}). As a result, $Q(r)$ (or the
``entropy'' in the case of a collisional fluid) shows a characteristic
upturn at $r_{\rm shock}$ like the one shown in
Fig.~\ref{FigSimSolProf}.

As discussed by \citet{White1993}, the first caustic/shock, which
occurs at about a third of the current turnaround radius, lies close
to the virial radius, as defined here. Given that our halos have
concentrations, $c=r_{200}/r_{-2}$, of order $7$-$10$
\citep{Neto2007}, we would then expect the upturn in the $Q$ profiles
to occur roughly at $\sim 7$-$10\, r_{-2}$.

The dashed black curves in Fig.~\ref{FigSimSolProf} show the
similarity solution, assuming $r_{\rm shock}=8 \, r_{-2}$, and
normalized vertically at $r=r_{-2}$ to coincide with the simulated
halo profiles. It is clear from Fig.~\ref{FigSimSolProf} that the
similarity solution is in excellent agreement with the $Q$ profiles of
our simulated halos. This suggests that the upturn in the outer $Q$
profiles just reflects the fact that the regions around the virial
radius have not yet fully virialized.

\section{Summary}
\label{SecSumm} 

We have used a set of 21 high-resolution cosmological N-body
simulations to investigate the pseudo-phase-space density profiles,
$Q(r)=\rho/\sigma^3$, of cold dark matter halos. In particular, we
concentrate our analysis on the radial dependence of $Q$ for particles
in the smooth main halo component, after carefully removing
substructures and correcting for halo shape.

Our main result is that although $Q(r)$ is remarkably
well-approximated by a power law, a slight but systematic upturn from
the power-law profile is clearly seen in the outer regions of all our
simulated halos. Both the exponent of the power law, as well as the
upturn in the outer regions are consistent with the particular
secondary infall similarity solutions derived by
\citet{Bertschinger1985}.  In these solutions, the power-law inner
region corresponds to the virialized region of the halo, whereas the
upturn in the outer regions coincides with the location of the
shock/first caustic of the system or, roughly speaking, with the
virial boundary of a halo. Although Bertschinger's is just one
particular secondary-infall similarity solution, valid in the simple
case of a point-mass perturber in an Einstein-de Sitter universe, the
upturn in Q marking the virial boundary of the system is expected to
be a general feature of such solutions.

Although the outer upturn may be robust, the fact that the exponent of
the inner power law is consistent with Bertschinger's ($\chi \sim
-1.875$) is somewhat surprising.  As shown by
\citet[][]{Fillmore1984}, the non-linear structure of halos
formed through self-similar secondary infall depend on the scaling
index that characterizes the mass dependence of the initial
perturbation, $\delta M/M \propto M^{-\epsilon}$. In the case of
Bertschinger's solution $\epsilon=1$, but this is a poor approximation to the
typical overdensity that seeds the collapse of galaxy-sized LCDM
halos. The agreement between Bertschinger's $Q(r)$ and the simulated
profiles is therefore a non-trivial result whose significance remains
unclear.



The presence of the upturn in the $Q(r)$ profiles casts doubts on work
that attempts to construct dynamical equilibrium models of CDM halos
by assuming that the power-law behaviour of $Q$ profiles applies to
{\it all radii} \citep[see, e.g.,][]{Austin2005,Dehnen2005}. The
$r^{\chi}$ behaviour seems to hold only in the inner regions, and our
results caution against fitting power laws to $Q(r)$ over a radial
range that extends outside the scale radius.

Because the radius where the upturn becomes noticeable marks the
transition to the region where virial equilibrium no longer holds,
this radius, expressed as a fraction of the virial radius, may vary
systematically with halo mass, collapse time, or cosmological
parameters.  Indeed, the virial radius definition we adopt here does
not depend on the formation history of each object, but the boundary
of the region where virial equilibrium might hold does. For example, the
shock radius of an early collapsing halo that has accreted little mass
in the recent past might occur further away (in units of $r_{200}$)
than another system of similar mass that has assembled a considerable
fraction of its present-day mass more recently. It may be worth
checking whether the trends of $\chi$ with power spectrum and mode of
assembly noted in the literature \citep[see,
e.g.,][]{Knollmann2008,Wang2009} might be explained by this process.

Finally, we note that our highest resolution halos (those from the
Aquarius project) also present evidence of departures from a simple
power-law $Q$ profile near the innermost resolved radius. At the
moment it is unclear whether this indicates that the power-law
behaviour of $Q(r)$ is just an approximation that breaks down inside
some characteristic radius, or whether estimates of $Q(r)$ at those
radii might be affected by numerical uncertainties. For example, it is
easy to demonstrate that an isotropic halo whose mass profile follows
strictly an Einasto law cannot have a power-law $Q(r)$ law that
extends all the way to the centre \citep[see, e.g.,][]{Ma2009}. 
The radial dependence of these $Q(r)$ profiles may be more accurately 
represented by an Einasto-like form where the power-law index changes 
gradually but smoothly with radius. We plan to address this and other 
pertinent issues in forthcoming work.

\begin{center}
\begin{table*}
  \caption{
    We list, for each halo in our sample, a label, the
    Plummer-equivalent gravitational softening and convergence radius, as
    well as the virial radius at $z=0$. The structural parameters
    $V_{\rm max}$ and $r_{\rm max}$ identify the peak of the circular
    velocity curve for each halo. Finally, $M_{200}$ is the virial
    mass and $N_{200}$ is the number of particles within the virial
    radius. The $z_{\rm rel}$ column specifies, for non-Aquarius
    halos, the redshift of the most recent snapshot when the halo
    satisfies the relaxation criteria introduced in Sec.~\ref{SecRelCrit}.}

\begin{tabular}{c c c c c c c c c c c c}\hline \hline
Halo       &$\epsilon_G$&  $r_{\rm conv}$    &$r_{200}$&$r_{\rm max}$&$V_{\rm max}$&$M_{200}$&$N_{200
}$& $z_{\rm rel}$ \\
           &   [kpc/h]  &[kpc/h]& [kpc/h] & [kpc/h]      & [km/s]     & $[10^{10} M_{\odot}/h]$  & $[10^6]$& \\ \hline

Aq-A-2    & 4.8$\times 10^{-2}$ &  0.250 & 179.5 & 20.5  & 208.5& 134.5 & 134.5 &   -  \\ 
Aq-B-2    & 4.8$\times 10^{-2}$ & 0.219  & 137.0 & 29.3  & 157.7& 59.8  & 127.1 &   -  \\ 
Aq-C-2    & 4.8$\times 10^{-2}$ & 0.248  & 177.3 & 23.7  & 222.4& 129.5 & 126.8 &   -  \\ 
Aq-D-2    & 4.8$\times 10^{-2}$ & 0.281  & 177.3 & 39.5  & 203.2& 129.5 & 127.0 &   -  \\ 
Aq-E-2    & 4.8$\times 10^{-2}$ & 0.223  & 155.0 & 40.5  & 179.0& 86.5  & 123.6 &   -  \\ 
Aq-F-2    & 4.8$\times 10^{-2}$ & 0.209  & 152.7 & 31.2  & 169.1& 82.8  & 167.5 &   -  \\ \hline

h1    &  0.39      &    1.090      &  134.4 &  44.0  &  151.9  & 56.4    & 2.04 & 0.198 \\
h2    &  0.25      &    0.852      &  144.6 &  35.1  &  159.9  & 70.3    & 4.91 & 0.000 \\
h3    &  0.38      &    1.063      &  154.1 &  33.7  &  178.6  & 85.0    & 2.60 & 0.049 \\
h4    &  0.31      &    0.899      &  154.7 &  30.3  &  175.8  & 86.1    & 4.17 & 0.876 \\
h5    &  0.24      &    0.797      &  156.1 &  32.0  &  174.8  & 88.5    & 6.10 & 0.000 \\
h6    &  0.26      &    0.829      &  158.0 &  47.3  &  171.6  & 91.7    & 6.00 & 0.000 \\
h7    &  0.35      &    1.001      &  158.3 &  37.4  &  184.8  & 92.2    & 3.25 & 0.062 \\
h8    &  0.39      &    1.148      &  175.6 &  39.1  &  203.5  & 125.9   & 3.30 & 0.350 \\
h9    &  0.45      &    1.351      &  177.8 &  45.1  &  200.2  & 130.8   & 2.48 & 0.000 \\
h10   &  0.33      &    0.944      &  183.7 &  21.3  &  209.2  & 144.1   & 4.79 & 0.350 \\
h11   &  1.06      &    3.239      &  275.5 &  188.2 & 285.6   & 486.4   & 1.08 & 0.309 \\
h12   &  1.39      &    4.006      &  391.0 &  114.2 & 425.6   & 1389.5  & 1.26 & 0.140 \\
h13   &  1.61      &    4.418      &  396.1 &  86.9  &  422.9  & 1445.2  & 0.98 & 0.140 \\
h14   &  2.08      &    6.886      &  856.0 &  263.6 & 888.2   & 14581.8 & 2.62 & 0.030 \\
h15   &  3.65      &    11.595     &  981.6 &  444.8 & 1043.0  & 21993.1 & 1.16 & 0.094 \\

\hline
\end{tabular}
\label{TabNumSims}
\end{table*}
\end{center}

\begin{center}
\begin{table*}
  \caption{Values of the exponent $\chi$ obtained from $r^{\chi}$ fits
    to various $Q$ profiles, as indicated in the legend. $\chi_r$ refers
    to fits to the ``radial'' $Q_r$ profiles. The average $\chi$ for all
    halos is listed in the next-to-last row, together with its standard deviation . The average figure-of-merit for all halos,
    $\langle \psi_{\rm min}\rangle$, is listed in the last row of the
    table. }
\begin{tabular}{ccccccccccccccc|} \hline \hline
   & \multicolumn{7}{c|}{Spherically averaged: $Q(r)$, $Q_r(r)$} &
  \multicolumn{2}{c|}{Median: $Q^{\rm s}_{\rm i}(r)$} \\

Halo &  $\chi$ &&  $\chi_{\rm r}$ &&   $\chi$ &&  $\chi_{\rm r}$ &  $\chi$ &   $\chi$   \\ 

& \multicolumn{3}{c|}{($r_{\rm conv} < r < r_{-2}$)} && \multicolumn{3}{c|}{($r_{\rm conv} < r < r_{200}$)} & \multicolumn{1}{l|}{($r_{\rm conv} < r < r_{-2}$)} & \multicolumn{1}{l|}{
($r_{\rm conv} < r < r_{200}$)} \\

\hline  
Aq-A-2     & -1.917 && -1.976 && -1.873 && -1.955 & -1.910 &     -1.821  \\ 
Aq-B-2     & -1.868 && -1.938 &&  -1.832 && -1.872 & -1.914 &     -1.808  \\ 
Aq-C-2     & -1.948 && -2.010 && -1.883 && -1.948 & -1.932 &     -1.823  \\ 
Aq-D-2     & -1.862 && -1.942 && -1.831 && -1.901 & -1.939 &     -1.804  \\ 
Aq-E-2     & -1.912 && -1.947 &&  -1.894 && -1.902 & -1.863 &     -1.853  \\ 
Aq-F-2     & -1.911 && -1.980 &&  -1.885 && -1.954 & -1.965 &     -1.823  \\ \hline

h1         & -1.826 && -1.847 && -1.858 && -1.813 & -1.798 & -1.792 \\ 
h2         & -1.862 && -1.920 && -1.811 && -1.908 & -1.828 & -1.696 \\ 
h3         & -1.993 && -2.076 && -1.864 && -1.903 & -1.967 & -1.800 \\
h4         & -1.895 && -1.942 && -1.863 && -1.943 & -1.879 & -1.766 \\
h5         & -1.916 && -1.999 && -1.872 && -1.938 & -1.887 & -1.771 \\
h6         & -1.898 && -2.010 && -1.863 && -1.908 & -1.856 & -1.801 \\
h7         & -1.960 && -1.918 && -1.940 && -1.957 & -1.878 & -1.863 \\
h8         & -1.864 && -1.951 && -1.808 && -1.916 & -1.871 & -1.727 \\
h9         & -1.869 && -1.953 && -1.858 && -1.926 & -1.846 & -1.792 \\
h10        & -1.975 &&-2.042 && -1.871 && -1.991 & -1.972 & -1.754 \\
h11        & -1.896 && -2.008 && -1.843 && -1.845 & -1.903 & -1.756 \\
h12        & -1.879 && -1.949 && -1.875 && -2.023 & -1.864 & -1.758 \\
h13        & -1.843 && -1.955 && -1.847 && -1.882 & -1.768 & -1.724 \\
h14        & -1.810 && -1.902 && -1.817 && -1.886 & -1.801 & -1.732 \\
h15        & -1.847 && -1.966 && -1.835 && -1.919 & -1.765 & -1.648 \\\hline
$\langle\chi \rangle$  & $-1.893\pm 0.048$ && $-1.963 \pm 0.050$ && $-1.858\pm
0.031$ && $-1.919 \pm 0.047$ & $-1.876\pm 0.061$ & $-1.777 \pm 0.052$ \\
$\langle\psi_{\rm min} \rangle$   & $4.3 (\pm 1.3) \times 10^{-2}$ &&
$5.38 (\pm 0.16)\times 10^{-1}$ && $8.46 (\pm 2.1)\times 10^{-2}$ &&
$1.42 (\pm 0.50)\times 10^{-1}$ & $4.23 (\pm 1.79)\times 10^{-2}$ &
$1.18 (\pm 0.43)\times 10^{-1}$ \\

\end{tabular}
\label{TabChiFits}
\end{table*}
\end{center} 


{}

\bibliography{master}

\end{document}